\documentclass[conference,10pt]{IEEEtran}
\IEEEoverridecommandlockouts
\usepackage{algorithm}
\usepackage[noend]{algpseudocode}
\usepackage{graphicx}
\usepackage{subcaption}
\usepackage{amsmath}
\usepackage{amsfonts}
\usepackage{hyperref}
\usepackage{cleveref}
\usepackage[dvipsnames]{xcolor}
\makeatletter
\newcommand\fs@spaceruled{\def\@fs@cfont{\bfseries}\let\@fs@capt\floatc@ruled
  \def\@fs@pre{\vspace{.9\baselineskip}\hrule height2.8pt depth0pt \kern2pt}%
  \def\@fs@post{\kern5pt\hrule\relax}%
  \def\@fs@mid{\kern5pt\hrule\kern2pt}%
  \let\@fs@iftopcapt\iftrue}
\makeatother

\begin{document}


\title{Large-Scale Spectrum Occupancy Learning via Tensor Decomposition and LSTM  Networks}

\author{\IEEEauthorblockN{Mohsen Joneidi$^*$, Ismail Alkhouri$^*$, and Nazanin~Rahnavard
\IEEEauthorblockA{{Department of Electrical and Computer Engineering}\\
{University of Central Florida}\\
Email: \{ialkhouri@knights., mohsen.joneidi@, nazanin@eecs.\}ucf.edu}
\thanks{* Indicates shared first authorship. This material is based upon work supported by the National Science Foundation under Grant No. CCF-1718195.}
}
\vspace{-8mm}
}



\maketitle 

\begin{abstract}
A new paradigm for large-scale spectrum occupancy learning based on long short-term memory (LSTM) recurrent neural networks is proposed. 
Studies have shown that spectrum usage is a highly correlated time series. Moreover, there is a correlation for occupancy of spectrum between different frequency channels. Therefore, revealing all these correlations using learning and prediction of one-dimensional time series is not a trivial task. In this paper, we introduce a new framework for representing the spectrum measurements in a tensor format. Next, a time-series prediction method based on CANDECOMP/PARFAC (CP) tensor decomposition and LSTM recurrent neural networks is proposed.  The proposed method is computationally efficient and is able to capture different types of correlation within the measured spectrum. Moreover, it is robust against noise and missing entries of sensed spectrum. The superiority of the proposed method is evaluated over a large-scale synthetic dataset in terms of prediction accuracy and computational efficiency.  
\end{abstract}


\IEEEpeerreviewmaketitle

\section{Introduction}

Spectrum occupancy learning (SOL) aims to extract spectrum usage patterns at each frequency band over time. The learned model of spectrum occupancy facilitates the functionality of dynamic spectrum access. Spectrum sensing, optimal channel selection for opportunistic spectrum access, and resource allocation are some tasks that can be performed more efficiently by the prediction of spectrum usage~\cite{li2017intelligent}.\par
The SOL problem can be regarded as time series learning and prediction and its performance mainly depends on the underlying model for the time series analysis. Many statistical models and methods for spectrum usage prediction have been proposed in the last decade \cite{zhang2016usefulness}. Auto-regressive models, Markov models \cite{eltom2016hmm,wang2015analysis} and neural networks \cite{bai2015spectrum,eltholth2016spectrum} are exploited as the core models for spectrum time-series prediction. However, spectrum usage is a non-stationary process whose characteristics are time-dependent \cite{d2009distribution}. Other factors such as users' mobility and diverse demands of users make this process more complex. To overcome this challenging problem, deep learning methods are successfully implemented for capturing spectral usage patterns \cite{zhang2018citywide,yu2017spectrum}. 
Long short-term memory (LSTM)  and convolutional neural networks (CNNs) are  popular models for learning deep networks in various applications such as computer vision and pattern recognition problems \cite{yogatama2017generative, fan2016video}. 
However, these methods are still challenging for large-scale learning  of spectrum time series. The correlation of spectrum occupancy w.r.t time could be within a very large range. For example, averaged spectrum occupancy may correlate to that of one hour ago, but some network activities are daily or weekly \cite{willkomm2008primary}. Thus, spectrum occupancy at one time could be related to spectrum occupancy of one day or a week ago as well. Likewise, there might exist some spectrum patterns even in a larger scale over time. While conventional time-series prediction methods fail to reveal correlations in large lags, LSTM is able to capture these patterns. However, there are two issues in the large-scale data scenario. First, learning and prediction of an extremely long time series implies capturing all the spectrum correlations efficiently and the computational burden of learning and updating the LSTM model may not be tractable for online tracking of spectrum occupancy. Second, dealing with missing entries in the learning phase is inevitable for a real data sequence as it affects the prediction accuracy in the test phase. We propose to utilize tensor-based data completion methods that had attracted many attentions for data processing in the presence of missing entries \cite{li2019tensor}.\par

This paper proposes a new high-dimensional structure for sensed spectrum data in order to improve accuracy and scalability of LSTM for large-scale SOL. A joint problem of data interpolation and extrapolation (completion and prediction) is introduced. Tensor CP decomposition provides a reliable low-dimensional representation of data, and LSTM performs a fast prediction on the lower-dimension data (decomposed factors).\par

The correlations in the matrix-based representation with a long lag are vulnerable to be forgotten. However, these correlations can be identified in a much smaller lag in the third dimension of a tensor. In the present paper, tensor-based representation of time series is exploited in order to extract some basic time series known as CP factors of a tensor. These factors are robust against noise and missing entries. Large-scale prediction of all time series over long-time dimension only requires a prediction of CP factors of the measured tensor. This significant advantage can be considered as a big data reduction technique.\par

The main contributions of this paper are summarized as following:
\begin{itemize}
\item A novel time-series prediction framework is proposed based on tensor decomposition and LSTM networks. Our framework can be employed for large-scale spectrum occupancy learning among many other large-scale time-series prediction applications.
\item Computational burden is decreased by solely performing prediction on low-dimensional CP factors rather than high-dimensional raw data. 
\item The problem of missing samples in the time-series prediction is addressed using tensor completion techniques.
\end{itemize}

Throughout this paper, $\mathcal{X}$ denotes a three-way tensor, $\boldsymbol{X}$ denotes a matrix. Mode-n fiber of a tensor is a  vector obtained by fixing all modes except the $n^{\text{th}}$ mode and Mode-n matricized version of tensor is denoted by $\boldsymbol{X}_{(n)}$. $\boldsymbol{x}$ and $x$ represent a vector and a scalar, respectively. Hadamard product, outer product, and Khatri-rao product are  denoted as $*$, $\circ$, and $\odot$, respectively \cite{kolda2009tensor}. 





\section{Background and System Model}
In this section, the prerequisite background is presented, then the system model for spectrum aggregation is explained. 
\subsection{Tensor CP Decomposition}
A Tensor is a multi-dimensional array. Since their introduction,
tensors have been utilized in various fields
as they bring a concise mathematical framework for formulating
and challenging problems involving high-dimensional
data or big data especially in signal processing \cite{sidiropoulos2017tensor}. The CP decomposition factorizes a 3-dimensional tensor $\mathcal{X}\in \mathbb{R}^ {F\times T\times N}$ of rank $R$ into a sum of rank-1 tensors which can be represented as  \cite{kolda2009tensor} 

\begin{equation}\label{CPD_ALS}
\small
{\mathcal{X}} = \; \sum_{r=1}^R  \mathbf{a}_r\circ \mathbf{b}_r\circ  \mathbf{c}_r\;\;\overset{\Delta}{=}<\boldsymbol{A},\boldsymbol{B},\boldsymbol{C}>,
\end{equation}

\normalsize{where, $\mathbf{a}_r$, $\mathbf{b}_r$, and $\mathbf{c}_r$ are the CP factors of the $r^{\text{th}}$ component and the $r^{\text{th}}$ column of factor matrices
$\boldsymbol{A}$, $\boldsymbol{B}$ and $\boldsymbol{C}$, respectively. In other words,
$\boldsymbol{A}= [\boldsymbol{a}_1\;\boldsymbol{a}_2\;\ldots \boldsymbol{a}_r]\in \mathbb{R}^{F\times R}$. 
Similarly, $\boldsymbol{B}\in \mathbb{R}^{T\times R}$ and $\boldsymbol{C}\in \mathbb{R}^{N\times R}$} are defined.

The tensor $\mathcal{X}$ can be matricized as \cite{kolda2009tensor},
\small{
\begin{equation}\label{CPD_ALS mat}
\mathbf{X}_{(1)} =\mathbf{A}(\mathbf{C}\odot \mathbf{B})^{T},\nonumber
\end{equation}
\begin{equation}\label{CPD_ALS mat}
\mathbf{X}_{(2)} =\mathbf{B}(\mathbf{C}\odot \mathbf{A})^{T},
\end{equation}
\begin{equation}\label{CPD_ALS mat}
\mathbf{X}_{(3)} =\mathbf{C}(\mathbf{B}\odot \mathbf{A})^{T}.\nonumber
\end{equation}
}
\normalsize{
A powerful property of high-order tensors is that their rank decomposition is unique under milder conditions compared to matrices \cite{sidiropoulos2000uniqueness}. The interesting characteristics of tensors have attracted researchers in communication systems for channel estimation and blind coding in MIMO systems  \cite{qian2018tensor,da2018tensor}. \par
}
Computing the CPD used for this paper is done by the alternating least squares (ALS) method proposed by Carrol, and Harshman \cite{carroll1970analysis,harshman1970foundations}. The goal is to calculate a CPD with $R$ components that best approximate $\mathcal{X}$, i.e., to obtain

\small{
\begin{equation}\label{CPD-ALS opt.}
\underset{\hat{\mathcal{X}}}{\text{min}} \|\mathcal{X}-\hat{\mathcal{X}}|| \quad \text{s.t.} \; \; \hat{{\mathcal{X}}} = \sum_{r=1}^R= \mathbf{a}_r\circ \mathbf{b}_r\circ  \mathbf{c}_r.
\end{equation}
}
\normalsize{
The ALS algorithm fixes $\mathbf{B}$ and $\mathbf{C}$ to solve for $\mathbf{A}$, then fixes $\mathbf{A}$ and $\mathbf{C}$ to solve for $\mathbf{B}$, and then fixes $\mathbf{A}$ and $\mathbf{B}$ to solve for $\mathbf{C}$ \cite{kolda2009tensor}. We refer to this algorithm as the plain CP algorithm.\par
}

\vspace{-1mm}
\subsection{LSTM Network}
Neural networks have been recognized as powerful techniques for spectrum pattern learning \cite{azmat2016analysis}. Similar to other neural network structures, LSTM consists of an input layer, hidden layer(s), and an output layer. LSTM network was introduced by Hochreiter and Schmidhuber in 1997 as an advanced type of recurrent neural networks (RNNs) \cite{hochreiter1997long}. RNN (or vanilla RNN) provides the feature of internal memory maintenance i.e., it saves the information of the previous time step. This method introduced the problem of gradient explosion which means that the network overwrites its memory in an uncontrolled manner. The main advantage of LSTM is to fix the issue encountered in the conventional RNNs by adding an adaptive memory unit, which is its key component.  This adaptive memory unit controls saving dominant samples and/or forgetting obsolete data. This feature enables LSTM to track information over longer periods of time. The mathematical computation of one memory cell is given in \cite{hochreiter1997long}.\par


\begin{figure}[t]
\centering
\includegraphics[width=8.4cm,height=5.5cm]{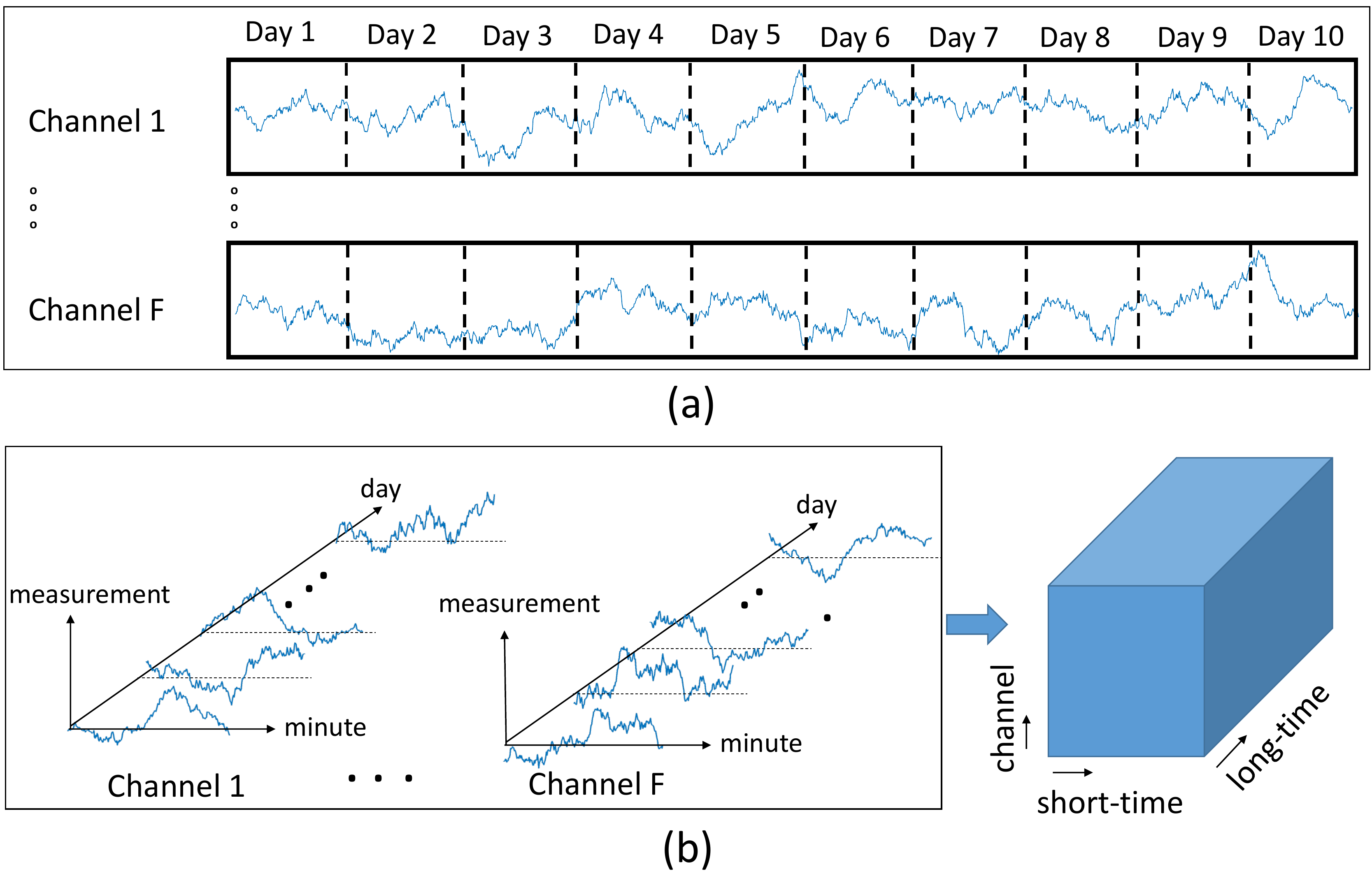}
\vspace{-1mm}
\caption{\small{Comparing two representing methods of  $F$  time series. (a) Matrix-based representation. (b) Tensor-based representation.}}
\vspace{-3mm}
\label{fig:data_structure}
\end{figure}

\subsection{System model}
Consider that we have a frequency spectrum sensor, that saves the power spectral density (PSD) of an RF band of $F$ frequency bins at $T$ times a day. Therefore, we will have a matrix of size ${T\times F}$ for each day of recording. Additionally, assume data is recorded for $N$ days. Corresponding to one frequency bin $f$, there exists a matrix of length $T\times N$ that presents occupancy changes over all times. Similarly, corresponding to one instant of time $t$, a matrix of length $F\times N$ represents the occupancy of spectrum over all channels and days in a specific time of day. After $N$ number of days (number of frontal slices), the occupancy of the ${f}^{th}$ frequency channel at ${t}^{th}$ time slot is the subject of prediction for the upcoming day. The proposed data arrangement is shown in Fig.\ref{fig:data_structure}.

To forecast the values of next time steps of a sequence, we utilize a  predictor that trains a regression network. Four types of training networks are used in this paper,  the auto-regressive model (AR), support vector machines (SVM), convolutional neural networks (CNN), and LSTM.\par

\section{Proposed Method}

The measured tensor consists of $TF$ number of time series that has $N$ values over long time (days) each. Prediction of every time series independently comes with two issues; (i) The long lag correlation between time series is neglected, thus noise and missing entries can easily affect prediction severely. (ii) Prediction of each time series implies learning a network which requires a large amount of computational burden.\par

Tensor decomposition learns a few principle factors for each way, such that all fibers  in the corresponding way of the tensor can be reconstructed using the linear combination of the learned fibers. Tensor CP representation is a concise model, and it is robust against perturbations and missing entries. It is shown that a low-rank tensor can be recovered from a small number of entries using CP decomposition. In other words, the CP factors of the original tensor and the CP factors of the partial and noisy replica of the original tensor are close to each other \cite{wang2018noisy}. These attractive characteristics of structured data in a high-dimensional tensor motivate us to employ tensor CP decomposition for dynamic spectrum completion and prediction. 

\begin{figure}[h]
\centering
\vspace{-4mm}
\includegraphics[width=7.5cm]{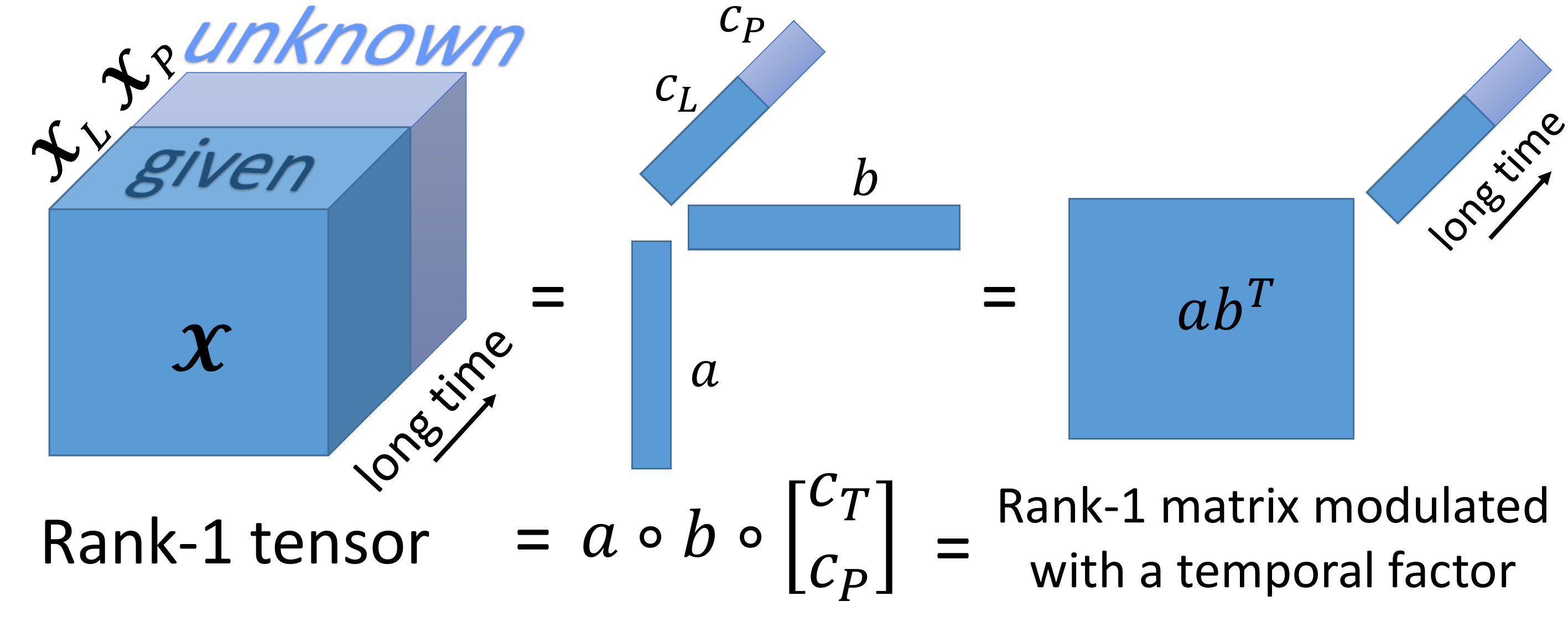}
\caption{\small{A rank-1 tensor is the outer product of $3$ vectors and it can be cast as the modulation of a rank-1 matrix with a temporal pattern.}}
\vspace{-2mm}
\label{fig:inution1}
\end{figure}

\begin{figure}[h]
\centering
\includegraphics[width=6.2cm]{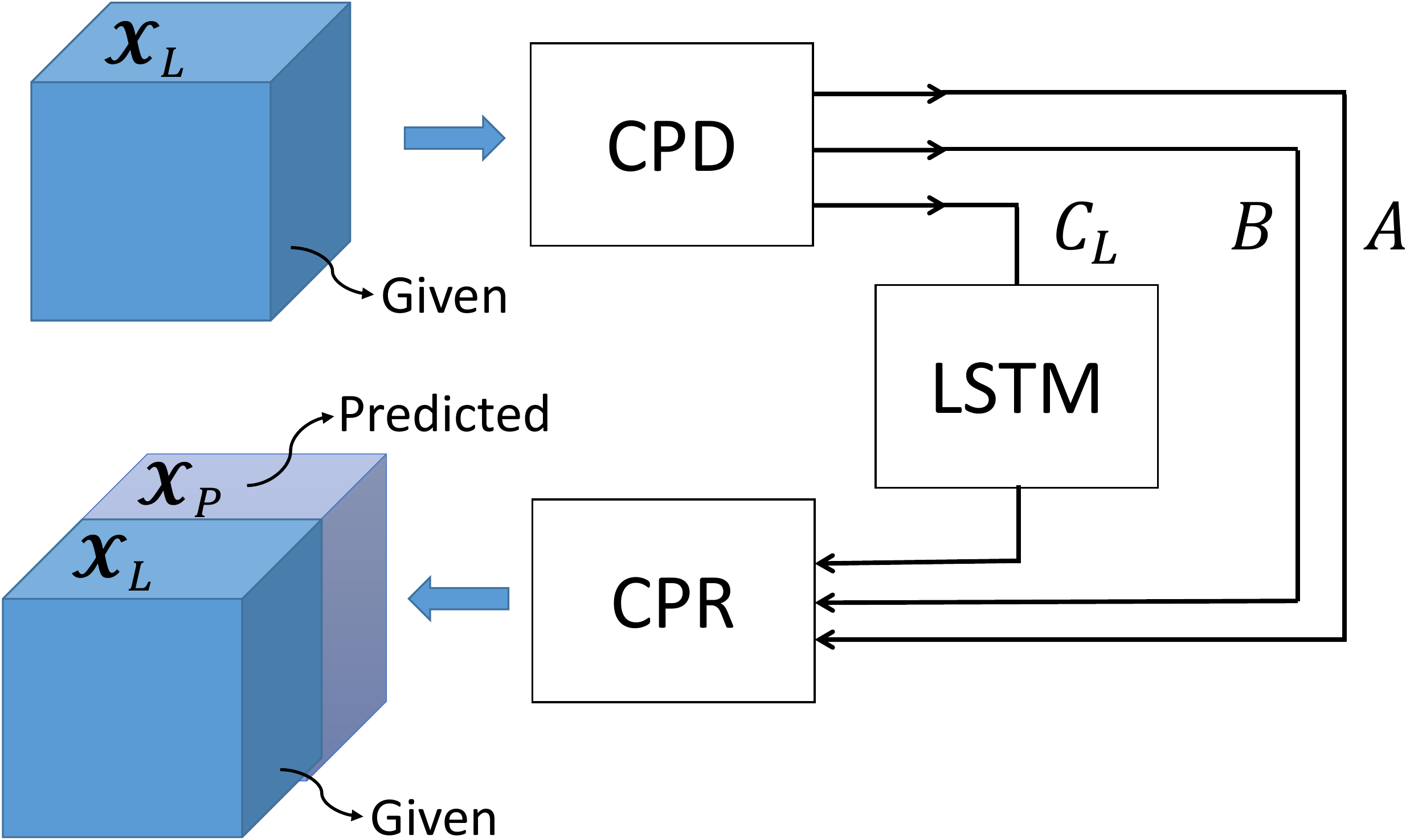}
\caption{\small{Block diagram of our tensor-based time-series prediction.}}
\vspace{-2mm}
\label{fig:block}
\end{figure}

Consider a rank-1 tensor $\boldsymbol{\mathcal{X}}\in \mathbb{R}^{F\times T \times N}$ with a third dimension over long time variable $n$ in Fig. \ref{fig:inution1}. 
This tensor consists of $FT$ time series (fibers) alongside its third way. Because $\boldsymbol{\mathcal{X}}$ has rank 1, all these time series are 
a scale of a vector $\boldsymbol{c}$, which is a basic time series.
This vector is broken down into two parts, the given part, $\boldsymbol{c}_L$, which corresponds to the known part of tensor, and the unknown part, $\boldsymbol{c}_P$, which corresponds to the part of the tensor to be predicted. 
Prediction of all unknown variables of tensors is equivalent to prediction of $\boldsymbol{c}_P$. For a general rank-$R$ tensor, there exist $R$ basic time series that span the space of all fibers of the tensor in the third way. Thus, the prediction of $R$ temporal factors enables us to predict any time series of the tensor. 
Suppose our source of data is dynamic, therefore, obsolete data might degrade the result of prediction. To tackle this problem, only recent slices are considered for learning. The number of slices for each epoch of prediction is referred as the length of training, $N_L$. Likewise, we define the length of prediction, $N_P$, where $N=N_P+N_L$ is equal to the size of third dimension of the underlying tensor. The proposed tensor-based prediction solves two following consecutive problems to predict the unknown entries of the tensor over time:

\vspace{-2mm}
\begin{subequations}
\small
\begin{align}
(\boldsymbol{A},\boldsymbol{B},\boldsymbol{C}_L)&=\underset{\boldsymbol{A},\boldsymbol{B},\boldsymbol{C}_L}{\text{argmin}}\;\|\mathcal{X}_L - < \boldsymbol{ A},\boldsymbol{B},\boldsymbol{C}_L>\|_F^2,\\
\vspace{5mm}
\mathcal{X}_P&=\;\;<\boldsymbol{A},\boldsymbol{B},f(\boldsymbol{C}_L,\Omega)>,
\end{align}
\end{subequations}
in which, $f(.\;,.)$ represents a model for time-series prediction and $\Omega$ is the set of model's parameters. We will investigate the effect of the prediction model on the performance of the whole framework. AR, SVM, CNN, and LSTM are studied as core models for prediction. However, our main proposed algorithm is LSTM-based. Fig. \ref{fig:block} shows the block diagram of the proposed prediction algorithm.

CP decomposition reveals the latent factors of data from different perspective, and LSTM predicts the long temporal factors. The extracted factors using CP and extrapolated factors using LSTM can produce a tensor by CP reconstruction (CPR). 
\begin{algorithm}[b]
\small
\caption{Time-series completion and prediction via tensor CP decomposition and LSTM prediction.}
\textbf{Input}: Incomplete Tensor $\mathcal{X}_L^I$, mask $\mathcal{M}$.\\
\textbf{Output}: Completed and predicted tensor $\hat{\mathcal{X}}$\\
\vspace{1.2mm}
\hspace{-2.5mm} \small{1:}   \normalsize $\boldsymbol{A}, \boldsymbol{B}, \boldsymbol{C}_L \leftarrow$ CP decomposition on  $\mathcal{X}_L^I$. \\ \vspace{1.2mm}
\textbf{While} (The stopping criterion is not met)\\
 \vspace{1mm}
 \hspace{-1mm} \small{2:}  $\qquad \mathcal{X}_L \leftarrow$ using Eq. (\ref{eq:compl}).\\
  \vspace{2mm}
   \hspace{-1mm} \small{3:}  $\qquad \boldsymbol{A}, \boldsymbol{B}, \boldsymbol{C}_L \leftarrow$ CP decomposition on $\mathcal{X}_L$.\\
   \vspace{-2mm}
\hspace{1mm}\textbf{End}\\

\hspace{-1mm} \small{4:} $\boldsymbol{C}_P\;\leftarrow$LSTM on each column of $\boldsymbol{C}_L$\vspace{-1mm}\\

\hspace{-1mm} \small{5:} $\boldsymbol{C}\leftarrow$ concatenate  $\boldsymbol{C}_L$ and $\boldsymbol{C}_P$\vspace{-1mm}\\

\hspace{-1mm} \small{6:} $\hat{\mathcal{X}}\leftarrow\;\; <\boldsymbol{A},\boldsymbol{B},\boldsymbol{C}>$.
\label{alg:prediction}
\end{algorithm}

Since tensor analysis  considers multi-dimensional correlation of data, tensor completion is a state-of-the-art method for data completion in many applications \cite{montanari2018spectral,yang2018tensor}. The proposed tensor-based scheme can be extended to the joint completion and prediction in a straightforward formulation. Assume a given incomplete tensor, $\mathcal{X}_L^I$ and a mask tensor with the same size of the data tensor, $\mathcal{M}\in \{ 0,1\}^{F\times T \times N_L}$. The entries corresponding to $0$ are not measured. The incomplete tensor can be completed using the formula given by 
\begin{equation}
\small
\label{eq:compl}
\mathcal{X}_L=\mathcal{M}\ast \mathcal{X}_L^I +(\underline{\boldsymbol{1}}-\mathcal{M})*<\boldsymbol{A},\boldsymbol{B},\boldsymbol{C}_L>.
\end{equation}

In which, $\underline{\boldsymbol{1}}$ is a tensor with all entries equal to $1$. The given data is kept and the missed data is estimated using CP factors. However, updating the incomplete tensor enables the algorithm to estimate a more accurate set of factors. Thus, CP factors and missing entries can be updated iteratively.  Alg. \ref{alg:prediction} shows the proposed method for joint time series completion and prediction. The main loop of the algorithm completes data to find a more fitted set of CP factors. Then, LSTM predicts the long-time factors, and the predicted time series are resulted by CP reconstruction. The tensor completion is performed using iterative CP decomposition and data interpolation. However, the exploited CP does not use the information of mask, and the mask is used only for data interpolation in~(\ref{eq:compl}). A modified version of CP decomposition is presented in Alg. \ref{alg:OP_CP} that infuses the information of mask in order to estimate CP factors of an incomplete tensor. The optimized CP for incomplete data can be employed in line 3 of Alg. \ref{alg:prediction} instead of the plain CP in order to estimate more accurate factors. 

\begin{algorithm}[htbp]
\small
\caption{Optimized CP for incomplete data.}
\textbf{Input}: Tensor $\mathcal{X}$, mask $\mathcal{M}$, and rank, $R$.\\
\textbf{Output}: CP factors of $\mathcal{X}$\\
\vspace{1.2mm}
\hspace{-2mm} \small{1:}  $\boldsymbol{A}, \boldsymbol{B}, \boldsymbol{C} \leftarrow$ Plain CP decomposition on  $\mathcal{X}$ \cite{kolda2009tensor}. \\ \vspace{1.2mm}
\textbf{While} (The stopping criterion is not met)\\
 \vspace{1mm}
\small{2:} $\qquad \boldsymbol{A}\;\leftarrow\; \underset{\boldsymbol{A}}{\text{minimize}}\; \|\boldsymbol{M}_{(1)} * (\boldsymbol{X}_{(1)}-\boldsymbol{A}(\boldsymbol{C}\odot \boldsymbol{B})^T)\|_F^2$\\ \vspace{2mm}
   \small{3:}  $\qquad \boldsymbol{B}\;\leftarrow\; \underset{\boldsymbol{B}}{\text{minimize}}\; \|\boldsymbol{M}_{(2)} * (\boldsymbol{X}_{(2)}-\boldsymbol{B}(\boldsymbol{C}\odot \boldsymbol{A})^T)\|_F^2$\\ \vspace{2mm}
   \small{4:}   $\qquad \boldsymbol{C}\;\leftarrow\; \underset{\boldsymbol{C}}{\text{minimize}}\; \|\boldsymbol{M}_{(3)} * (\boldsymbol{X}_{(3)}-\boldsymbol{C}(\boldsymbol{B}\odot \boldsymbol{A})^T)\|_F^2$\\
$\;\;\quad$\textbf{End}
\label{alg:OP_CP}
\end{algorithm}

SOL can be regarded as a learning-based detection  where the problem is to detect whether a channel is occupied or not. Our decision rule for detection is based on the output of our proposed algorithm. Assume $\hat{x}_{ftn}$ is the predicted spectrum value at frequency channel $f$, time $t$ and day $n$. Two hypotheses are considered for spectrum occupancy status for this entry.
\vspace{-2mm}
\begin{equation}
\small
\label{eq:detection}
S(f,t,n) =
\left\{
	\begin{array}{ll}
		\text{OCCUPIED} & \mbox{if } \hat{x}_{ftn} \geq \gamma \\
		\text{NOT OCCUPIED} & \mbox{if } \hat{x}_{ftn} < \gamma
	\end{array}
\right.
\end{equation}

In which, $S(f,t,n)$ indicates the estimated occupancy status at frequency channel $f$, time $t$ and day $n$ and $\gamma$ is a threshold for operating the designed detector.  As $\gamma$ increases, both the probability of detection and the probability of false alarm will decrease. Receiver operating characteristic (ROC) of the proposed detector is able to find the optimum threshold to achieve the desired false alarm rate. 

\section{Experimental Results}

In the following experiments, we assume that 20 frequency channels are sensed. PSD of each frequency channel is recorded  10 times an hour, i.e., there exist 240 measurements from the spectrum for each mode-2 fiber. Moreover,  it is assumed that the recording for 100 days is available. Therefore, ${F}=20$, ${T}=240$, and ${N}=100$.\par

Synthetic dataset for time $t$ at day $n$ and frequency $f$ follows the joint probability distribution of $P(t,n,f)=P_t(t)P_n(n)P_f(f)$ where each distribution is generated according to the below model,

\begin{subequations}
\label{eq:dataset}
\small
\begin{equation}
P_t(t)= \sum_{i=1}^3 \beta_{i} \mathcal{N}(\tau_{i},\sigma_{i}^{2}),\qquad\qquad\quad\;
\end{equation}
\vspace{-3mm}
\begin{equation}
P_n(n|j) = \mathcal{N}(\mu_{j},\lambda_{j}^{2}), \;\text{for}\;\; j=n\;\text{mod}\;7,
\end{equation}
\vspace{-4mm}
\begin{equation}
P_f(f) =\mathcal{U}[1,2,...,F].\qquad\qquad \qquad \quad
\end{equation}
\end{subequations}

(\ref{eq:dataset}a) is the probability of spectrum occupation in a typical day which is modeled by a Gaussian mixture model (GMM) with three peaks at 3pm, 6pm, and 9pm. Parameters $\{\beta_i,\tau_i,\sigma_i\}$ are designed to satisfy the desired pattern of GMM\footnote{$\beta_1$=$0.5,\;\beta_2$=$0.3,\;\beta_3$=$0.2$,$\;\tau_1$=$150\;(3\text{PM}),\tau_2$=$180 \;(6\text{PM}),\tau_3$=$210\;$ $(9\text{PM})$, and $\sigma_i$=$20\;(2\text{hour})$}. The conditional probability of occupancy over days  follows (\ref{eq:dataset}b). The condition determines that $n$ corresponds to which day of week.  The parameters $\{\mu_j,\lambda_j\}$ are designed such that at Mondays, Tuesdays, Wednesdays, and Thursdays, the spectrum is more occupied than Fridays, and Friday is busier than the weekend\footnote{$\mu_0=\mu_1=\mu_2=\mu_3=1$, $\mu_4=0.5$,  $\mu_5=\mu_6=0.2$, and $\lambda_j=0.1\mu_j$} \cite{zhang2018citywide}.  In addition, there is no preference for frequency occupation of a user which leads to a uniform distribution with equal probabilities over all frequency bins, which is employed in (\ref{eq:dataset}c). This model is inferred from previous work \cite{willkomm2008primary}.

Selected LSTM parameters are 4 hidden layers with 4 unites each. Learning rate is 0.05 and the number of epochs is 300 with ADAM optimizer. Intel Corei7 CPU with 4.20GHz and 8 GB RAM is used for performing simulations on MATLAB 2018b.\par



The CPD-ALS algorithm determines the factors of the tensor numerically by solving alternating optimization problems. Calculating CP rank of a tensor is an NP-hard problem. However, it is upper bounded by the following inequality \cite{kolda2009tensor},
$$
\small
\text{Rank}(\mathcal{X})\le \text{min}(FT,FN,TN).
$$
A practical solution for finding rank is to start with a low number, compute the normalized reconstruction error, and increase it as needed. Normalized error is obtained as a function of rank as follows,

\begin{equation}\label{sensing}
\small
{e}_{cpd}(R) = \frac{||\mathcal{X} - \hat{\mathcal{X}}(R)||_{F}}{||\mathcal{X}||_{F}}.
\end{equation}

In which, ${||.||_{F}}$ denotes matrix Frobinious norm and $R$ takes values from 1 to a maximum rank and $\hat{\mathcal{X}(R)}$ is the rank-$R$ approximation of $\mathcal{X}$ optimized by a tensor decomposition algorithm. The goal is to select the lowest rank that approximates ${\mathcal{X}}$. The effect of rank for training the basic time series will be investigated later.  

In this experiment, results of the proposed method is exhibited. The synthesized data is organized into a $F\times T \times N$ tensor. In which  ${F}=20$ (20 frequency bins), ${T}=240$ (240 measurements per days), and ${N}=100$ (100 days). with rank 10, CP decomposition provides $\mathbf{A}\in \mathbb{R}^{20\times 10}$, $\mathbf{B}\in \mathbb{R}^{240\times 10}$, and $\mathbf{C}\in \mathbb{R}^{100\times 10}$. In order to evaluate prediction performance, the underlying tensor is broken into two tensors, (i) the learning tensor, $\mathcal{X}_L$, and  (ii)  the test tensor, $\mathcal{X}_P$, that is the subject of prediction. In this experiment $N_L=80$ days are used for learning and $N_P=20$ days are considered for prediction. \par

The obtained long-time CP factors, $\boldsymbol{C}_L\in \mathbb{R}^{80\times 10}$ are exploited to predict  $\boldsymbol{C}_P\in \mathbb{R}^{20\times 10}$. Each column of $\boldsymbol{C}_L$ is a pseudo-time series that is employed for prediction of $\boldsymbol{C}_P$ independently. Predicted values from AR, SVM, CNN, and LSTM training networks are computed. We also calculated the prediction of the matrix-based data using the aforementioned training methods to demonstrate the impact of utilizing CPD. Numerical comparison with other methods is presented in Table \ref{tbl_compare}. Tensor-based methods improve prediction accuracy as well as save computational burden.\par

 Employing LSTM for prediction of CP factors exhibits the best results, and it decreases computation cost comparison to the plain LSTM on the set of raw time series. The normalized error is computed using the following rule,\par
$$
\small
e_p=\frac{\sum (x_i-\hat{x}_i)^2}{x_i^2},
$$
where, $x_i$ and $\hat{x}_i$ are the actual and the predicted values in the time series.


\begin{table}[t]
\vspace{+2mm}
\caption{ \small{{Normalized Prediction Error and Processing Time (sec)}}}
\label{tbl_compare}\centering
 \begin{tabular}{||c c c c c||} 
 \hline
 Method & CPD time & Learning time& Total time&Error\%\\ [0.5ex] 
 \hline\hline
 AR \cite{eltholth2015forward} & N/A & 55.12 & 55.12 & 33.55\\ 
 \hline
 AR+CPD & 3.71 & 4.23 & 7.94&21.83 \\
 \hline
 SVM \cite{azmat2016analysis}& N/A & 1202.21 & 1202.21&23.78 \\

 \hline
 SVM+CPD & 3.71 & 20.52 & 24.23&16.94 \\
 \hline
  
 CNN \cite{zhang2018citywide}& N/A & 496.44 & 496.44&22.40 \\
 \hline
  CNN+CPD & 3.71 & 15.87 & 19.58& 17.81 \\ 

  \hline
 LSTM \cite{yu2018spectrum} & N/A & 2389.96 & 2389.96&23.71 \\ 
 \hline
  LSTM+CPD & 3.71 & 12.01 & 15.72&\textbf{15.26} \\ 
 \hline
\end{tabular}
\vspace{-3mm}
\end{table}

It can be observed that each prediction technique is improved by employing CPD. Our proposed method, LSTM+CPD, returns the best performance in terms of  the normalized prediction error. In general, LSTM  outperforms methods based on AR, SVM, or CNN \cite{eltholth2015forward, azmat2016analysis, zhang2018citywide}. It is worthwhile to notice that our proposed method predicts spectrum occupancy more accurately than performing LSTM on raw time series data \cite{yu2018spectrum}. On top of the enhanced prediction error, CPD achieves a massive data reduction. Table \ref{tbl_compare} demonstrates the processing time for each method and illustrates that exploiting CPD is able to diminish the total running time of prediction rigorously.\par

\begin{figure}[t]
\centering
\begin{subfigure}{0.23\textwidth}
\includegraphics[width=1.6 in ,height=3cm]{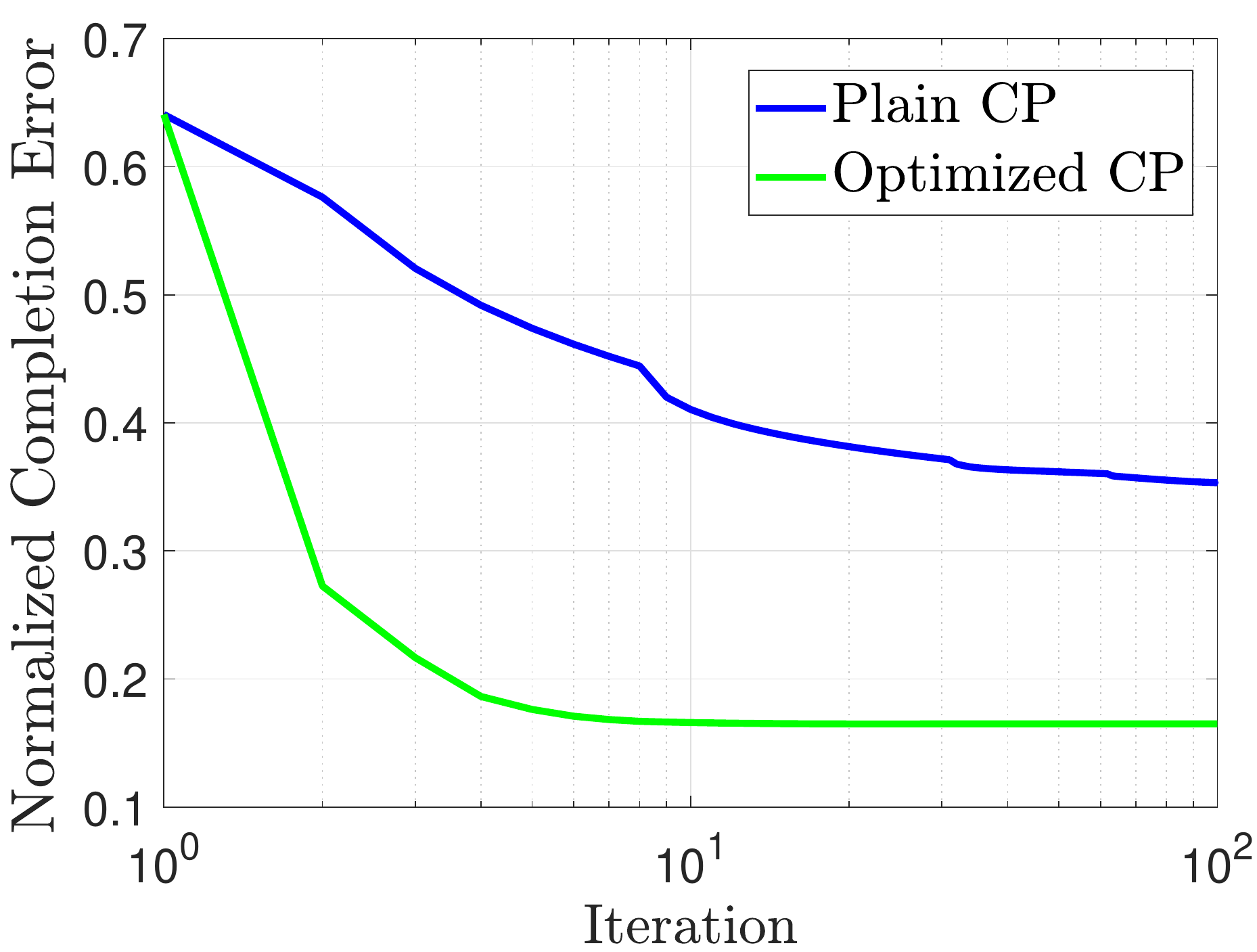}
\caption{ }
\end{subfigure}
\begin{subfigure}{0.23\textwidth}
\includegraphics[width=1.6 in,height=3cm]{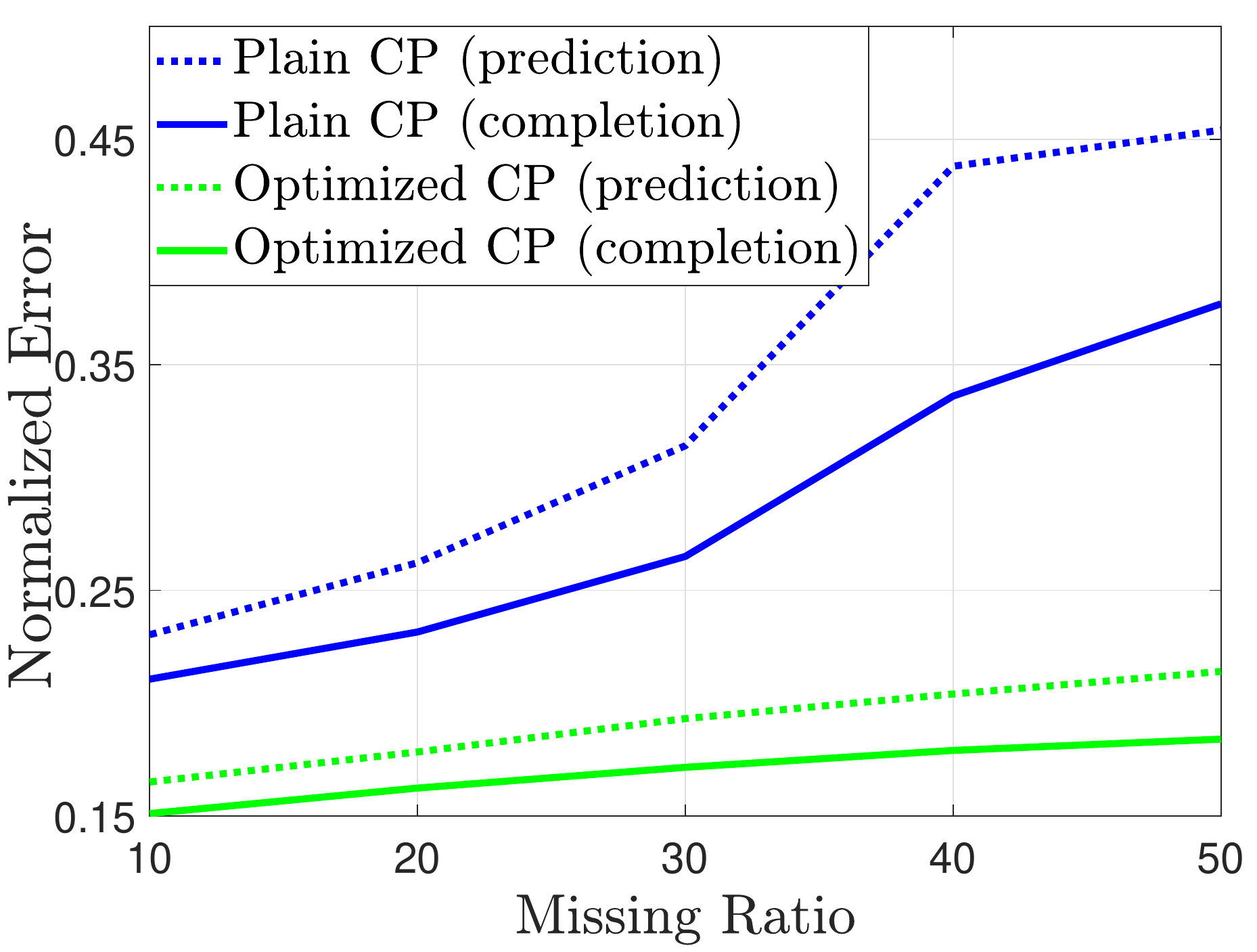}
\caption{}
\end{subfigure}
\caption{\small{Normalized completion error using the proposed method in Alg. \ref{alg:prediction} (a) Over iterations. (b) For different missing ratios.}}
\label{fig:compl}
\vspace{-3mm}
\end{figure}

In the next experiment the proposed method, Alg. \ref{alg:prediction}, is employed for missing spectrum recovery when a portion of spectrum measurements is missing. To this aim, the whole tensor is assumed to be incomplete. Therefore, random measurements from a $F\times T \times N$ tensor are available to recover the whole tensor.  The proposed spectrum  completion algorithm requires performing CPD in each iteration of completion. It is shown that employing the modified CP for incomplete data, Alg. \ref{alg:OP_CP}, is more effective for missing spectrum recovery.  Each iteration of data completion using the optimized CP needs more computation. however, the number of needed iterations for the modified CP is much less than the plain CP. Fig. \ref{fig:compl} (a) shows the performance of our proposed time series completion method using plain CP and the introduced CP versus iteration of data completion in Alg. \ref{alg:prediction}.\par

The performance of our proposed joint completion and prediction problem is presented for missing ratio in the range of $10$ and $50$ percent of data. Plain CP algorithm and the modified CP are compared for performing Alg. \ref{alg:prediction} to solve the joint problem. In this experiment, time series of $80$ days are considered for learning and $20$ days for prediction. The learning tensor, $\mathcal{X}_L$, is assumed to have missing entries. As it can be seen in Fig. \ref{fig:compl} (b), our proposed algorithm successfully completes data in terms of the normalized error and predicts time series using LSTM. As previously stated, the modified CP outperforms plain CP in presence of missing entries.  The prediction error is close to that of exploiting all data for learning that is presented in Table \ref{tbl_compare}. For example, in presence of $10\%$ missing entries for learning, the prediction error is $16.53\%$. This number is close to $15.26\%$  which is obtained by learning using the full tensor.





\begin{figure}[t]
\centering
\begin{subfigure}{0.23\textwidth}
\includegraphics[width=3.95cm,height=3cm]{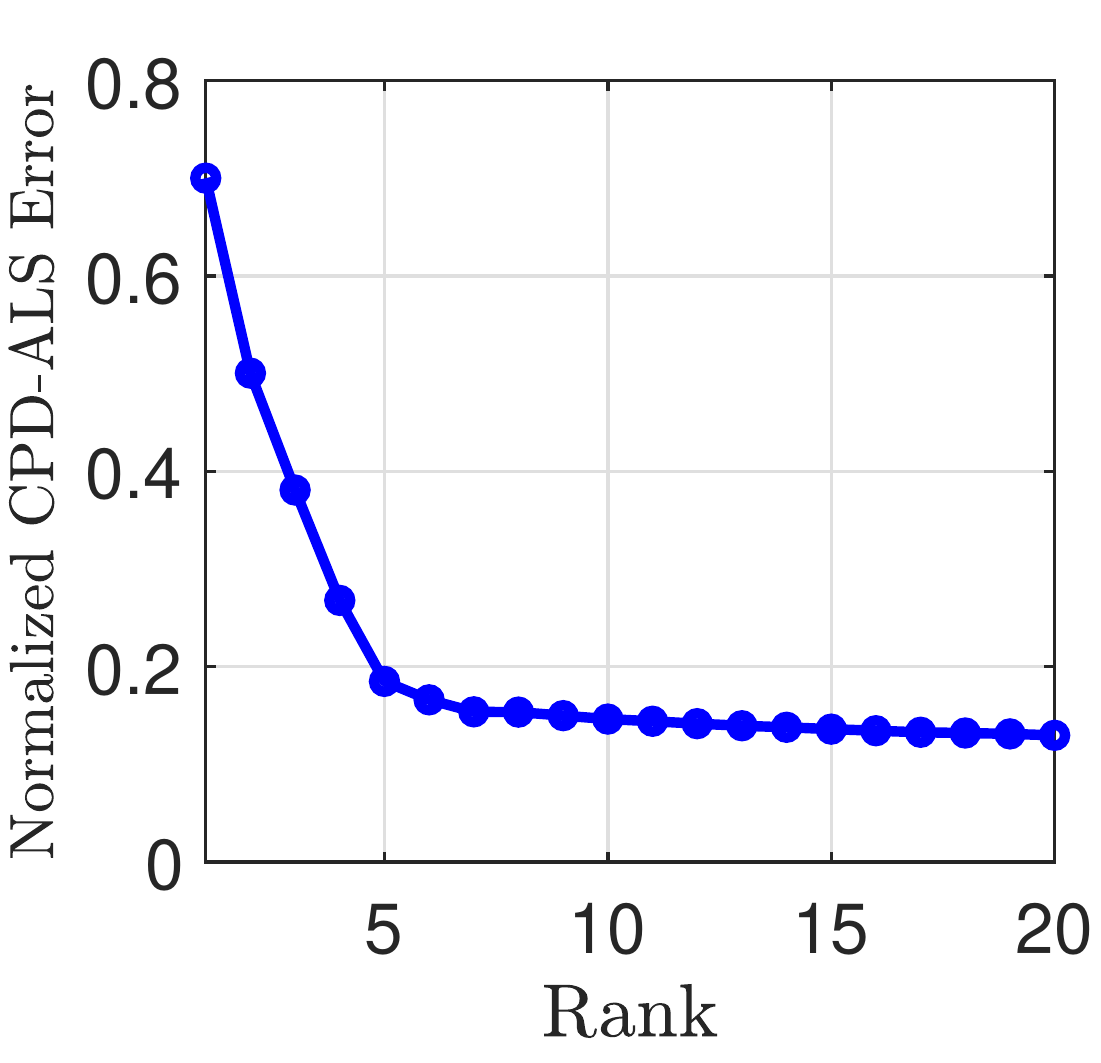}
\vspace{-2mm}
\caption{}
\end{subfigure}
\begin{subfigure}{0.23\textwidth}
\includegraphics[width=4.05cm,height=3cm]{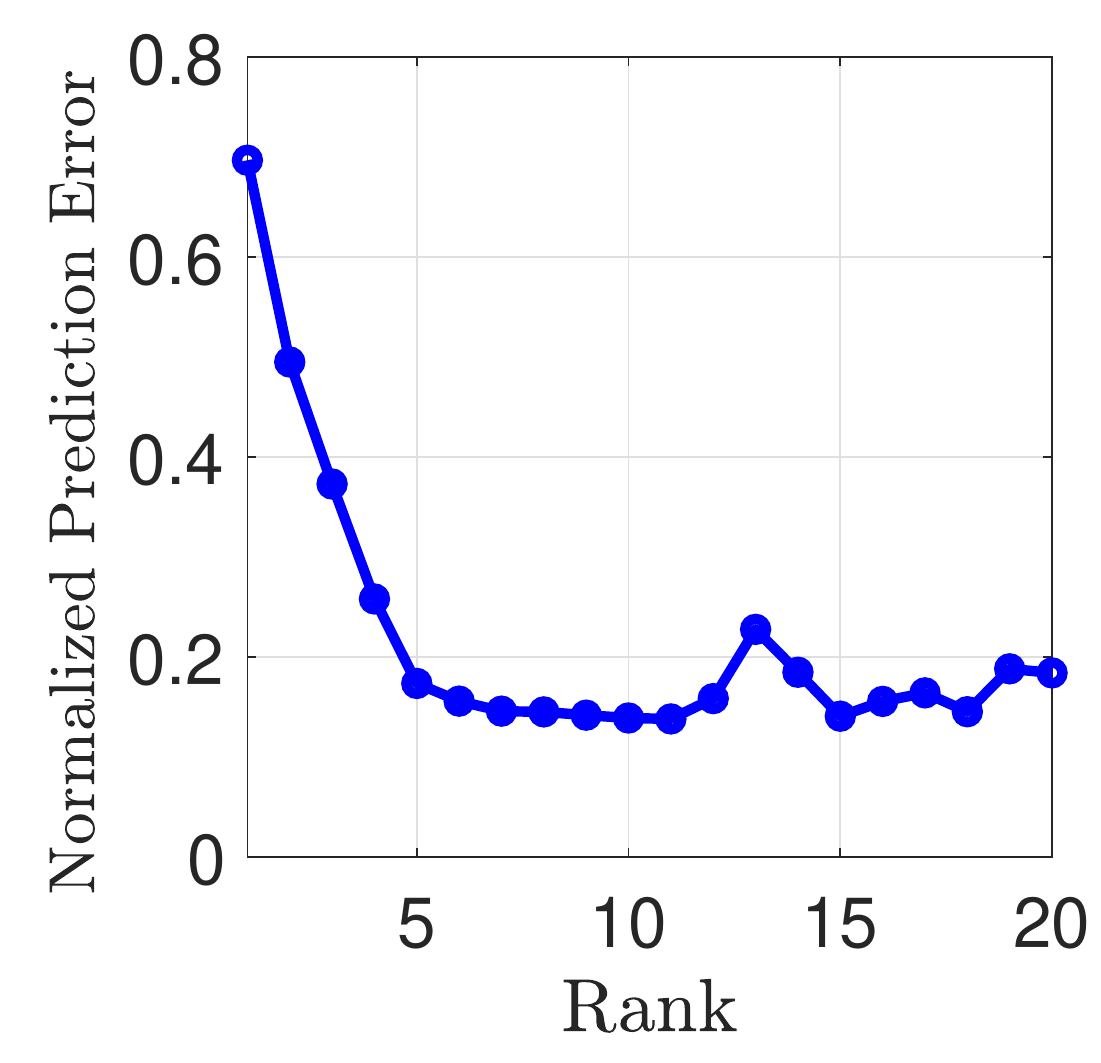}
\vspace{-2mm}
\caption{}
\end{subfigure}
\vspace{-2mm}
\caption{\small{Normalized error of prediction vs. the assumed rank for the underlying tensor. As the rank increases the learning error decreases. However, increasing rank causes over-learning for prediction. Thus, prediction error is not necessarily decreasing. }}
\vspace{-3mm}
\label{fig:rank_pred}
\end{figure}

Each component of CP decomposition learns some patterns of data. Selection of the rank equal to $R$ provides $R$ set of factors that reconstruct the learning tensor. As the assumed rank increases, more details about learning tensor are captured and the reconstruction error decreases. Fig. \ref{fig:rank_pred} (a) shows reconstruction error of the learning tensor versus the assumed rank.  However, learning fine details does not help prediction. Thus, the imposed rank can not be a large number arbitrarily. Fig. \ref{fig:rank_pred} (b) shows the performance of prediction using LSTM versus the selected rank of CP for decomposition of the learning tensor. As shown after rank 10, the normalized error of prediction is not decreasing by increasing rank.  

\begin{figure}[t]
\centering
\includegraphics[width=8.8cm,height=5.05cm]{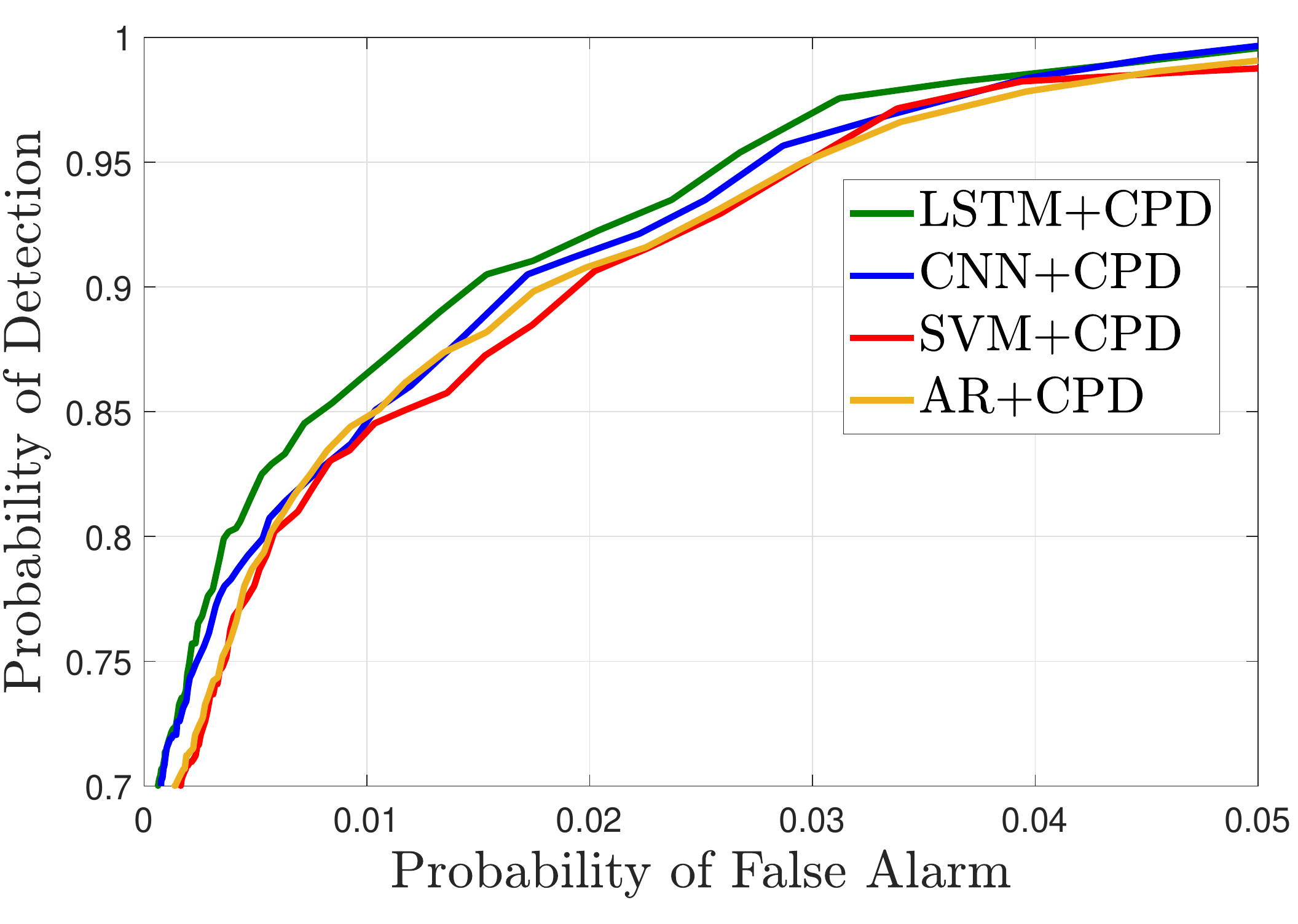}
\vspace{-5mm}
\caption{\small{{ROC of the proposed detector in Eq. \ref{eq:detection}.}}}
\label{fig:ROC}
\vspace{-3mm}
\end{figure}

 
The last experiment of this paper shows the performance of spectrum occupancy detection. Two hypotheses are considered based on Eq. (\ref{eq:detection}).The detection performance is determined using a ground truth of spectrum occupancy from the synthesized data. Our proposed spectrum prediction results in a value for spectrum in each channel over time. The value turns into a decision rule by Eq. (\ref{eq:detection}). Probability of detection, $P_D$, vs. probability of false alarm, $P_F$, are plotted by applying different values for the threshold. Utilization of AR, SVM, CNN, and LSTM on the tensor-based prediction is compared by their ROC graph in Fig. \ref{fig:ROC}. LSTM exhibits better performance for detection of free channels. It means that with a fixed false alarm rate, the  probability of detection using the proposed LSTM-based method is higher than the other methods.

\section{Conclusion}
In this paper, a combination of tensor decomposition and LSTM time-series prediction is proposed as a new paradigm for large-scale spectrum occupancy prediction. The measured spectrum data is  organized into a 3-way tensor. The CPD-ALS algorithm is performed to obtain CP factors for big data reduction and learning reliable patterns of data. The LSTM network is then utilized to predict CP factors in order to estimate future spectrum occupancy patterns over time and for all frequency channels. 
 Employing LSTM as the core predictor of CP factors outperforms other schemes such as AR, SVM, and CNN. The performance of handling missing data on the sensed spectrum illustrated robustness of CP factors against perturbations on the learning information. \par




\ifCLASSOPTIONcaptionsoff
  \newpage
\fi



%

 \bibliography{ref.bbl}
 \bibliographystyle{IEEEtran}




%








\end{document}